\documentclass{nature_preprint}

\usepackage[dvips]{graphics,color}

\title{Infrared radiation from an extrasolar planet}

\author{Drake Deming$^{1}$, Sara Seager$^{2}$ , L. Jeremy Richardson$^{3}$,
\& Joseph Harrington$^{4}$}

\begin{document}

\maketitle

\begin{affiliations}

\item Planetary Systems Laboratory and Goddard Center for
Astrobiology, Code 693, NASA's Goddard Space Flight Center,
Greenbelt, Maryland 20771, USA

\item Department of Terrestrial Magnetism, Carnegie Institution of 
Washington, 5241 Broad Branch Road NW, Washington, DC 20015, USA

\item Exoplanet and Stellar Astrophysics  Laboratory, Code 667, NASA's 
Goddard Space Flight Center, Greenbelt, Maryland 20771, USA

\item Center for Radiophysics and Space Research, Cornell University, 
326 Space Sciences Bldg, Ithaca, New York 14853-6801, USA

\end{affiliations}

\begin{abstract}

A class of extrasolar giant planets - the so-called `hot
Jupiters'\cite{1} - orbit within 0.05 AU of their primary stars. These
planets should be hot and so emit detectable infrared
radiation\cite{2}. The planet HD\,209458b\cite{3,4} is an ideal
candidate for the detection and characterization of this infrared
light because it is eclipsed by the star. This planet has an
anomalously large radius (1.35 times that of Jupiter\cite{5}), which
may be the result of ongoing tidal dissipation\cite{6}, but this
explanation requires a non-zero orbital eccentricity
($\sim$0.03)\cite{6,7}, maintained by interaction with a hypothetical
second planet. Here we report detection of infrared (24~$\mu$m)
radiation from HD\,209458b, by observing the decrement in flux during
secondary eclipse, when the planet passes behind the star. The
planet's 24~$\mu$m flux is 55$\pm$10 $\mu$Jy (1$\sigma$), with a
brightness temperature of 1130$\pm$150 Kelvins, confirming the
predicted heating by stellar irradiation\cite{2,8}. The secondary
eclipse occurs at the midpoint between transits of the planet in front
of the star (to within $\pm$7 min, 1$\sigma$), which means that a
dynamically significant orbital eccentricity is unlikely.

\end{abstract}

Operating cryogenically in a thermally stable space environment, the
Spitzer Space Telescope\cite{9} has sufficient sensitivity to detect
hot Jupiters at their predicted infrared flux levels\cite{8}. We
observed the secondary eclipse (hereafter referred to as `the
eclipse') of HD\,209458b with the 24~$\mu$m channel of the Multiband
Imaging Photometer for Spitzer (MIPS)\cite{10}. Our photometric time
series observations began on 6~December 2004 at 21:29 UTC, and ended
at approximately 03:23 UTC on 7~December 2004 (5 h 54 min
duration). We analyze 1696 of the 1728 10-s exposures so acquired,
rejecting 32 images having obvious flaws. The Supplementary
Information contains a sample image, together with information on the
noise properties of the data.

We first verify that circumstellar dust does not contribute
significantly to the stellar flux. Summing each stellar image over a
$13 \times 13$ pixel synthetic aperture ($33 \times 33$ arcsec), we
multiply the average sum by 1.15 to account for the far wings of the
point spread function (PSF)\cite{11}, deriving a flux of
21.17$\pm$0.11~mJy. The temperature of the star is close to
6000K\cite{12}. At a distance of 47 pc\cite{13}, a model
atmosphere\cite{14} predicts a flux of 22~mJy, agreeing with our
observed flux to within an estimated $\sim$~2~mJy error in absolute
calibration. We conclude that the observed flux is dominated by
photospheric emission, in agreement with a large Spitzer study of
planet-bearing stars at this wavelength\cite{11}.

Our time series analysis is optimized for high relative precision. We
extract the intensity of the star from each image using optimal
photometry with a spatial weighting function\cite{15}. Selecting the
Tiny Tim\cite{16} synthetic MIPS PSF for a 5000 Kelvin source at
24~$\mu$m, we spline-interpolate it to 0.01 pixel spacing, rebin it to
the data resolution, and center it on the stellar image. The best
centering is judged by a least-squares fit to the star, fitting to
within the noise level. The best-centered PSF becomes the weighting
function in deriving the stellar photometric intensity. We subtract
the average background over each image before applying the
weights. MIPS data includes per-pixel error estimates\cite{17}, which
we use in the optimal photometry and to compute errors for each
photometric point. The optimal algorithm\cite{15} predicts the
signal-to-noise ratio (SNR) for each photometric point, and these
average to 119. Our data are divided into 14 blocks, defined by
pre-determined raster positions of the star on the detector. To check
our SNR, we compute the internal scatter within each block. This gives
SNR in the range from 95 to 120 (averaging 111), in excellent
agreement with the optimal algorithm. For each point we use the most
conservative possible error: either the scatter within that block or
the algorithm estimate, whichever is greater. We search for
correlations between the photometric intensities and small
fluctuations in stellar position, but find none. We also perform
simple aperture photometry on the images, and this independent
procedure confirms our results, but with 60$\%$ greater errors.

The performance of MIPS at 24~$\mu$m is known to be
excellent\cite{18}. Only one instrument quirk affects our
photometry. The MIPS observing sequence obtains periodic bias images,
which reset the detector. Images following resets have lower overall
intensities (by $\sim$~0.1-1$\%$), which recover in later images. The
change is common to all pixels in the detector, and we remove it by
dividing the stellar intensities by the average zodiacal background in
each image. We thereby remove variations in instrument/detector
response, both known and unknown. The best available zodiacal
model\cite{19} predicts a background increase of 0.18$\%$ during the
$\sim$~6~h of our photometry. Because the star will not share this
increase, we remove a 0.18$\%$ linear baseline from the stellar
photometry. Note that the eclipse involves both a decrease and
increase in flux, and its detection is insensitive to monotonic linear
baseline effects.

Reliably detecting weak signals requires investigating the nature of
the errors. We find that shot noise in the zodiacal background is the
dominant source of error; systematic effects are undetectable after
normalizing any individual pixel to the total zodiacal background. All
of our results are based on analysis of the 1696 individual
photometric measurements versus heliocentric phase from a recent
ephemeris\cite{20} (Fig.~1a). We propagate the individual errors (not
shown on Fig.~1a) through a transit curve fit to calculate the error
on the eclipse depth. Because about half of the 1696 points are out of
eclipse, and half are in eclipse, and the SNR $\sim$ 111 per point,
the error on the eclipse depth should be $\sim(0.009)(2^{0.5})/(848^{0.5}) =
0.044\%$ of the stellar continuum. Model atmospheres for hot
Jupiters\cite{2,8,21,22,23,24} predict eclipse depths in the range from
0.2-0.4$\%$ of the stellar continuum, so we anticipate a detection of
4-9$\sigma$ significance. The eclipse is difficult to discern by eye
on Fig.~1a, because the observed depth (0.26$\%$) is a factor of 4
below the scatter of individual points. We use the known period (3.524
days) and radii\cite{5} to fit an eclipse curve to the Fig.~1a data,
varying only the eclipse depth, and constraining the central phase to
0.5. This fit detects the eclipse at a depth of 0.26$\%$ $\pm$
0.046$\%$, with a reduced ${\chi}^2$ of 0.963, denoting a good
fit. Note that the 5.6$\sigma$ significance applies to the aggregate
result, not to individual points. The eclipse is more readily seen by
eye on Fig.~1b, which presents binned data and the best-fit eclipse
curve. The data are divided into many bins, so the aggregate
5.6$\sigma$ significance is much less for a single bin (SNR $\sim$~1
per point). Nevertheless, the dip in flux due to the eclipse is
apparent, and the observed duration is approximately as expected. As a
check, we use a control photometric sequence (Fig.~1b) to eliminate
false positive detection of the eclipse due to instrument effects. We
also plot the distribution of points in intensity for both the
in-eclipse and out-of-eclipse phase intervals (Fig.~1c). This shows
that the entire distribution shifts as expected with the eclipse,
providing additional discrimination against a false positive
detection.

We further illustrate the reality of the eclipse on Fig.~2. Now
shifting the eclipse curve in phase, we find the best-fitting
amplitude and ${\chi}^2$ at each shift. This determines the best-fit
central phase for the eclipse, and also further illustrates the
statistical significance of the result. The thick line in Fig.~2a
shows that the maximum amplitude (0.26$\%$) is obtained at exactly
phase 0.5 (which is also the minimum of ${\chi}^2$). Further, we plot
the eclipse `amplitude' versus central phase using 100 sets of
synthetic data, consisting of gaussian noise with dispersion matching
the real data, but without an eclipse. The amplitude (0.26$\%$) of the
eclipse in the real data stands well above the statistical
fluctuations in the synthetic data.  Figure~2b shows confidence
intervals on the amplitude and central phase, based on the ${\chi}^2$
values. The phase shift of the eclipse is quite sensitive to
eccentricity ($e$) and is given\cite{25} as ${\delta}t = 2Pe
\cos(\omega)/\pi$, where $P$ is the orbital period, and $\omega$ is the
longitude of periastron. The Doppler data alone give $e=0.027 \pm
0.015$ (Laughlin, G., personal communication), and allow a phase shift
as large as $\pm 0.017$ (87 min). We find the eclipse centred at phase
0.5, and we checked the precision using a bootstrap Monte Carlo
procedure\cite{26}. The $1\sigma$ phase error from this method is
0.0015 ($\sim$~7~min), consistent with Fig.~2b. A dynamically
significant eccentricity, $e~\sim~0.03$\cite{6,7}, constrained by our
3$\sigma$ limit of ${\delta}t < 21$~min, requires $|({\omega} -
{\pi}/2)| <$ 12 degrees, and is therefore only possible in the unlikely
case that our viewing angle is closely parallel to the major axis of
the orbit. A circular orbit rules out a promising explanation for the
planet's anomalously large radius: tidal dissipation as an interior
energy source to slow down planetary evolution and
contraction\cite{7}. Because the dynamical time for tidal decay to a
circular orbit is short, this scenario posited the presence of a
perturbing second planet in the system to continually force the
eccentricity - a planet that is no longer necessary with a circular
orbit for HD\,209458b. The infrared flux from the planet follows
directly from our measured stellar flux (21.2~mJy) and the eclipse
depth (0.26$\%$), giving 55~$\pm$~10~$\mu$Jy. The error is dominated
by uncertainty in the eclipse depth. Using the planet's known
radius\cite{5} and distance\cite{13}, we obtain a brightness
temperature T$_{24}$=1130 $\pm$ 150K, confirming heating by stellar
irradiation\cite{2}. Nevertheless, T$_{24}$ could differ significantly
from the temperature of the equivalent blackbody (T$_{eq}$), that is,
one whose bolometric flux is the same as the planet. Without
measurements at shorter wavelengths, a model atmosphere must be used
to estimate T$_{eq}$ from the 24~$\mu$m flux. One such model is shown
in Fig.~3, having T$_{eq}$ =1700K. This temperature is much higher than
T$_{24}$ (1130K) due to strong, continuous H$_{2}$O vapor absorption at
24~$\mu$m. The bulk of the planetary thermal emission derives
ultimately from re-radiated stellar irradiation, and is emitted at
1-4~$\mu$m, between H$_{2}$O bands. However, our 24~$\mu$m flux error
admits a range of models, including some with a significantly lower
T$_{eq}$ (for example, but not limited to, models with reflective clouds
or less H$_{2}$O vapor).

Shortly after submission of this Letter, we became aware of a similar
detection for the TrES-1 transiting planet system\cite{27} using
Spitzer's Infrared Array Camera\cite{28}. Together, these Spitzer
results represent the first measurement of radiation from extrasolar
planets. Additional Spitzer observations should rapidly narrow the
range of acceptable models, and reveal the atmospheric structure,
composition, and other characteristics of close-in extrasolar giant
planets.
{~~~~}
{~~~~}





\begin{addendum}

\item  We thank G. Laughlin for communicating the
latest orbital eccentricity solutions from the Doppler data and for
his evaluation of their status. We acknowledge informative
conversations with D.~Charbonneau, G.~Marcy, B.~Hansen, K.~Menou and
J.~Cho. This work is based on observations made with the Spitzer Space
Telescope, which is operated by the Jet Propulsion Laboratory,
California Institute of Technology, under contract to NASA. Support
for this work was provided directly by NASA, and by its Origins of
Solar Systems program and Astrobiology Institute. We thank all the
personnel of the Spitzer telescope and the MIPS instrument, who
ultimately made these measurements possible. L.J.R. is a National
Research Council Associate at NASA's Goddard Space Flight Center.

\item[Competing Interests] The authors declare that they have no competing 
financial interests.

\item[Correspondence] and requests for materials should be addressed to 
D.D.~(email: Leo.D.Deming@nasa.gov).

\end{addendum}

\begin{figure}

\caption{Observations showing our detection of the secondary 
eclipse in HD\,209458. {\bf a}, Relative intensities versus
heliocentric phase (scale at top) for all 1696 data points. The phase
is corrected for light travel time at the orbital position of the
telescope. Error bars are suppressed for clarity. The gap in the data
near phase 0.497 is due to a pause for telescope overhead
activity. The secondary eclipse is present, but is a factor of
$\sim$~4 below the $\sim$~1$\%$ noise level of a single
measurement. {\bf b}, Intensities from a, averaged in bins of phase
width 0.001 (scale at top), with 1$\sigma$ error bars computed by
statistical combination from the errors of individual points. The red
line is the best-fit secondary eclipse curve (depth=0.26$\%$),
constrained to a central phase of 0.5. The points in blue are a
control sequence, summing intensities over a $10 \times 10$ pixel
region of the detector, to beat down the random errors and reveal any
possible systematic effects. The control sequence uses the same
detector pixels, on average, as those where the star resides, but is
sampled out of phase with the variations in the star's raster motion
during the MIPS photometry cycle. {\bf c}, Histograms of intensity
(lower abscissa scale) for the points in {\bf a}, with bin width
0.1$\%$, shown separately for the out-of-eclipse (black) and
in-eclipse intervals (red).}

\end{figure}

\begin{figure}
\caption{ Amplitude of the secondary eclipse versus assumed central 
phase, with confidence intervals for both. {\bf a}, The darkest line
shows the amplitude of the best-fit eclipse curve versus the assumed
central phase (scale at top). The overplotted point marks the fit
having smallest ${\chi}^2$, which also has the greatest eclipse
amplitude. The numerous thinner black lines show the effect of fitting
to 100 synthetic data sets containing no eclipse, but having the same
per-point errors as the real data. Their fluctuations in retrieved
amplitude versus phase are indicative of the error in eclipse
amplitude, and are consistent with $\sigma=$0.046$\%$. Note that the
eclipse amplitude found in the real data (0.26$\%$) stands well above
the error envelope at phase 0.5. {\bf b}, Confidence intervals at the
1, 2, 3 and 4$\sigma$ levels for the eclipse amplitude and central
phase (note expansion of phase scale, at bottom). The plotted point
marks the best fit (minimum ${\chi}^2$) with eclipse depth of
0.26$\%$, and central phase indistinguishable from 0.5. The center of
the eclipse occurs in our data at Julian day 2453346.5278.}

\end{figure}

\begin{figure}
\caption{Flux from a model atmosphere shown in
comparison to our measured infrared flux at 24~$\mu$m. A theoretical
spectrum (solid line) shows that planetary emission (dominated by
absorbed and re-radiated stellar radiation) should be very different
from a blackbody. Hence, models are required to interpret the
24~$\mu$m flux measurement in terms of the planetary temperature. The
model shown has an equivalent blackbody temperature T$_{eq}$ =1700 K
and was computed from a one-dimensional plane-parallel radiative
transfer model, considering a solar system abundance of gases, no
clouds, and the absorbed stellar radiation re-emitted on the day side
only. Note the marked difference from a 1700K blackbody (dashed line),
although the total flux integrated over the blackbody spectrum is
equal to the total flux integrated over the model spectrum. (The peaks
at short wavelength dominate the flux integral in the atmosphere
model, note log scale in the ordinate.) The suppressed flux at
24~$\mu$m is due to water vapor opacity. This model lies at the hot
end of the range of plausible models consistent with our measurement,
but the error bars admit models with cooler T$_{eq}$. }
\end{figure}






\clearpage
\includegraphics[110,160][700,750]{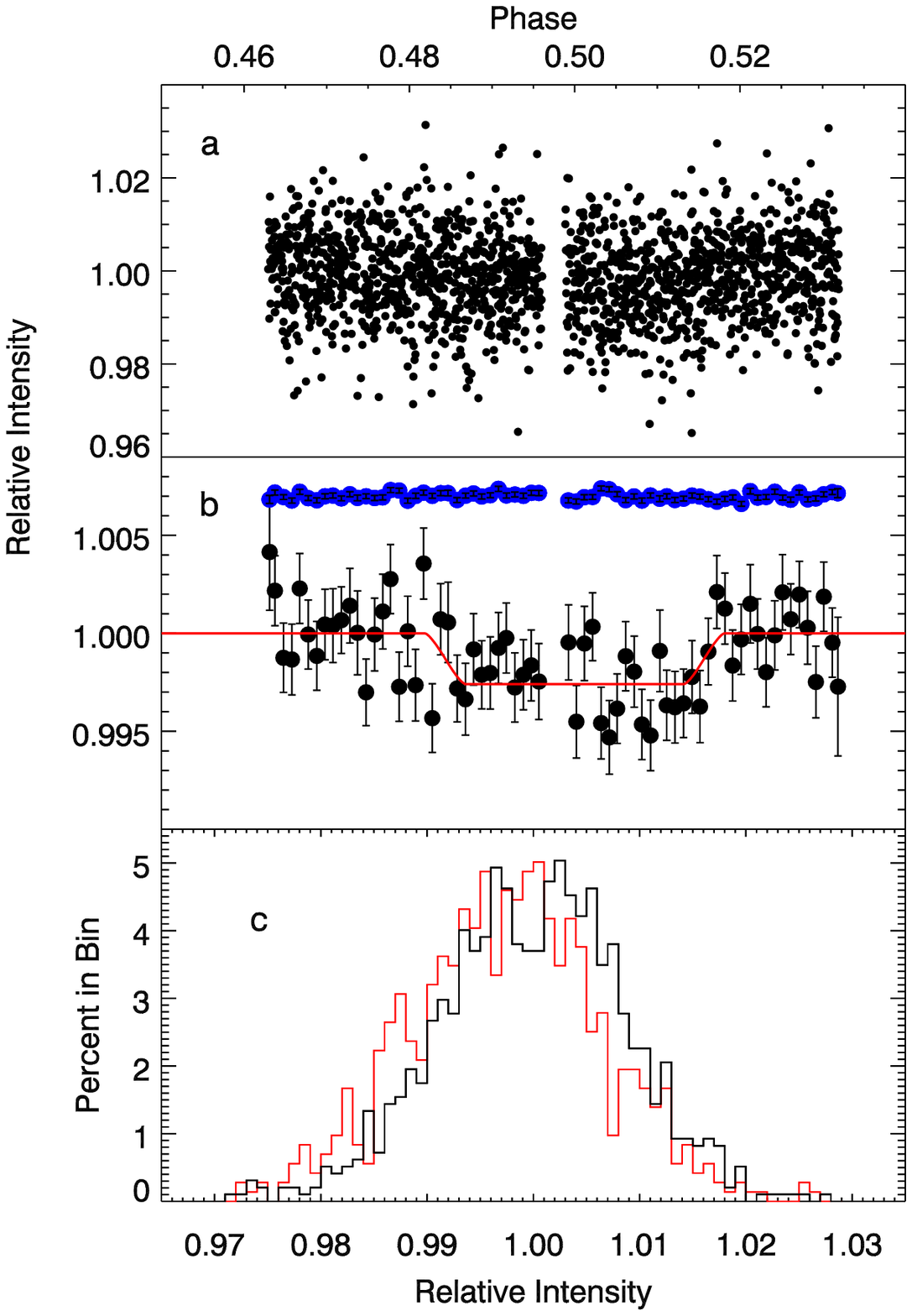}
\clearpage
\includegraphics[100,300][480,750]{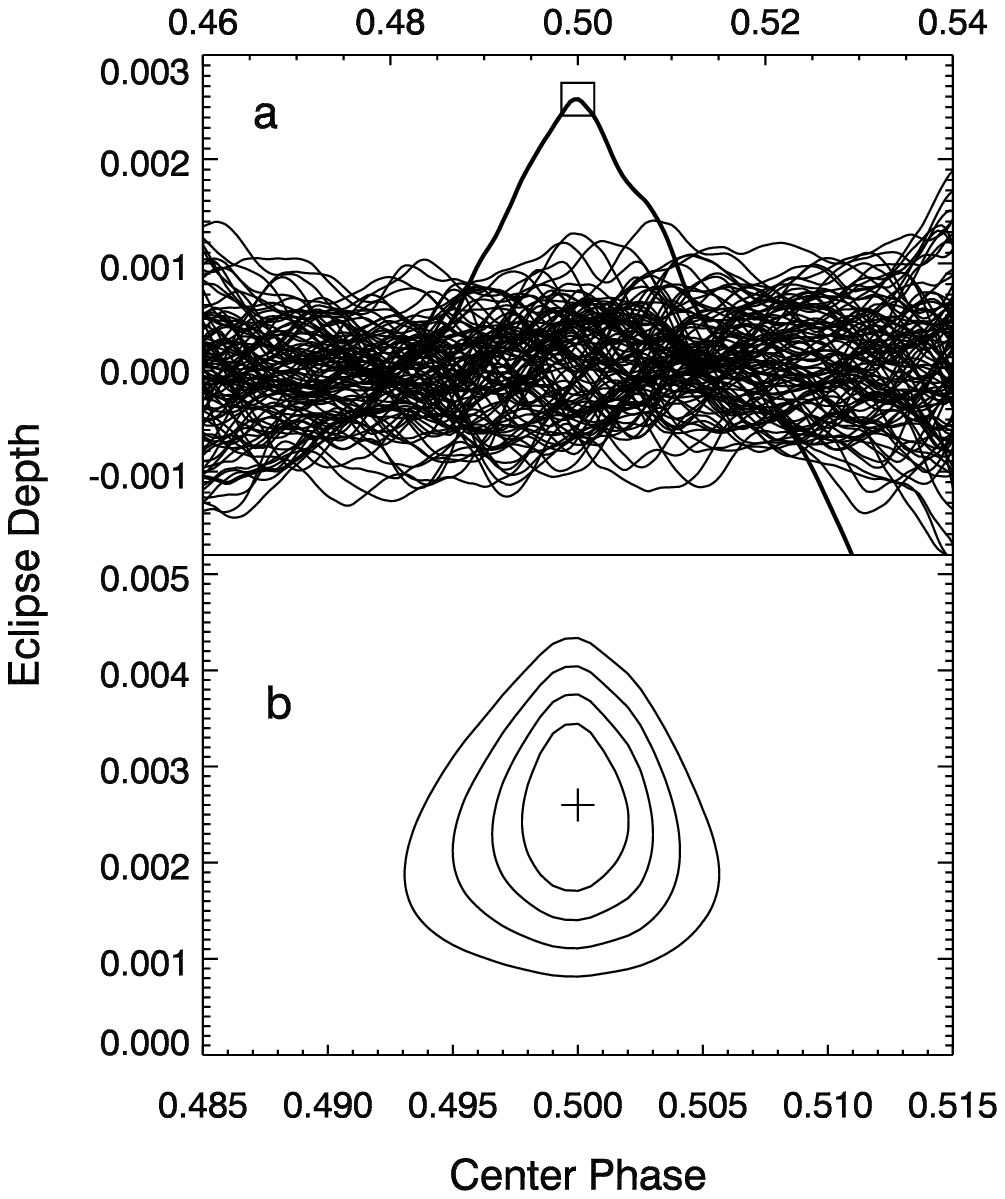}
\clearpage
\includegraphics[110,200][500,750]{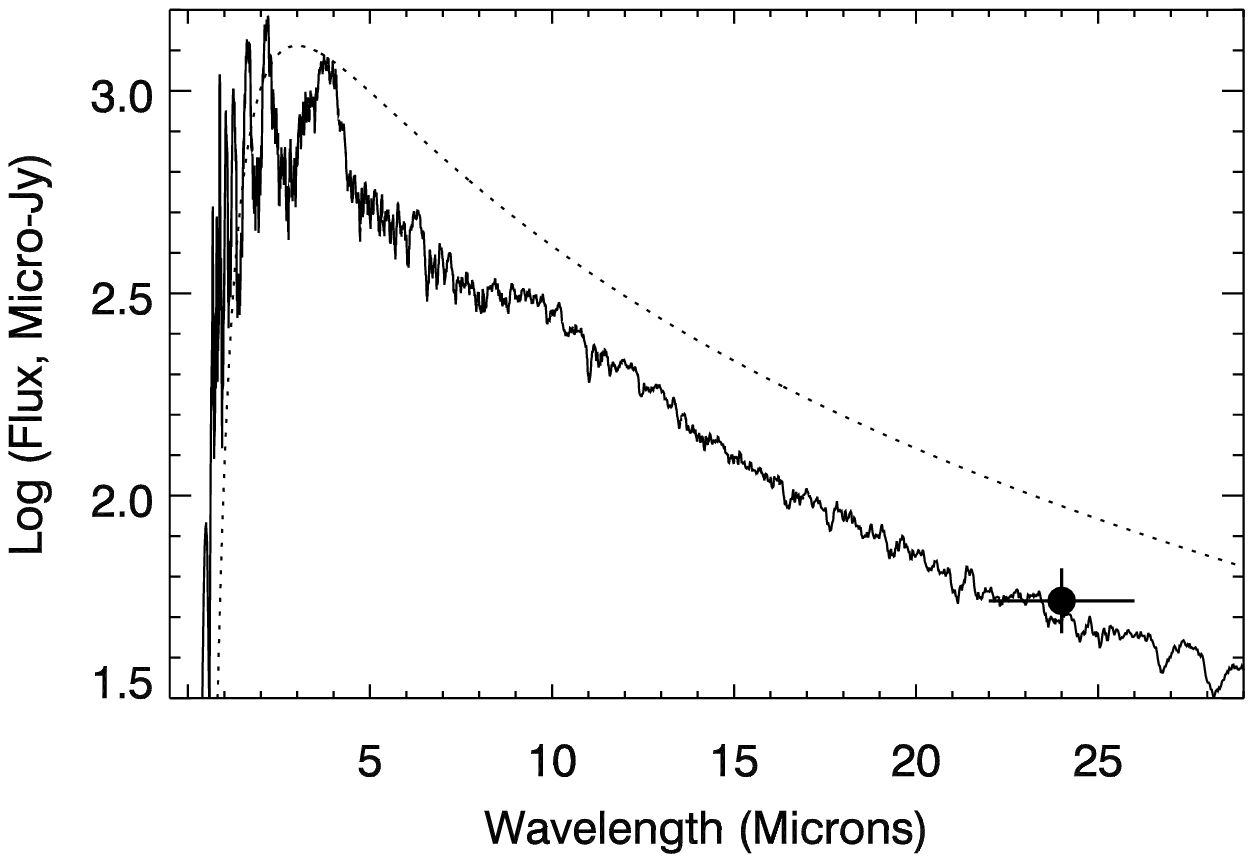}

\end{document}